# An interesting result concerning the lower bound to the energy in the Heisenberg picture.

Dan Solomon
Rauland-Borg Corporation
3450 W Oakton
Skokie, IL 60076  USA

Email: dan.solomon@rauland.com
August 18, 2008

**Abstract.**

In quantum theory it is generally assumed that there exists a special state called the vacuum state and that this state is a lower bound to the energy. However it has recently been demonstrated that this is not necessarily the case for some situations [5]. In order clarify the situation we will consider a "very simple" field theory in the Heisenberg picture consisting of a quantized fermion field with zero mass particles in 1-1D space-time interacting with a classical electrical potential. It will be shown that for this example there is no lower bound to the energy.

## 1. Introduction.

In quantum theory it is generally assumed that there exists a special state called the vacuum state and that this state is a lower bound to the energy. That is, no state can have a value of energy that this less than that of the vacuum state. However it has been shown by the author that this is not necessarily the case for some situations. For example it can be easily shown that in Dirac's hole theory there exist states with less energy than that of the vacuum state [1][2][3][4]. It has also been recently demonstrated that for quantum field theory in the Heisenberg representation there are states with less energy than the vacuum [5].

In order to clarify the situation we examine a "very simple" field theory in the Heisenberg picture. The field theory will consist of a quantized fermion field consisting of non-interacting fermions with zero mass. This fermion field will interact with a classical potential in 1-1D space-time. The advantage of this formulation is that it is possible to obtain exact solutions to the equations of motion. It will be shown that for this field theory there is no lower bound to the energy.

In the Heisenberg picture the state vectors $|\Omega\rangle$ are constant in time and the time dependence of the quantum state is carried by the field operators $\hat{\psi}(z,t)$ where $z$ is used to represent the space dimension. This is in contrast to the Schrödinger picture where the field operators are constant in time and the time dependence goes with the state vectors. Both pictures are presumed to be equivalent, however this assumption has been challenged by P.A.M. Dirac [6][7]. Some differences between the two pictures are also discussed [8]. In the rest of this paper we will focus solely on the Heisenberg picture.

## 2. The Heisenberg picture.

As was stated in the Introduction we will assume that the electrons have zero mass and are non-interacting, i.e., they only interact with an external electric potential. In addition we will work in 1-1 dimensional space-time where the space dimension is taken along the z-axis and use natural units so that $\hbar = c = 1$. This allows us to simplify the discussion and avoid unnecessary mathematical details. In this formulation an exact solution to the equations of motion is readily achieved as will be shown in the following discussion.

In the Heisenberg picture the field operators evolve in time according to the Dirac equation (see Chapt. 9 of [9] or Section 8 of [10] or Ref. [5]). For 1-1D space time the Dirac equation can be written as,

$$i\frac{\partial \hat{\psi}(z,t)}{\partial t} = H\hat{\psi}(z,t) \tag{2.1}$$

where the Dirac Hamiltonian is given by,

$$H = H_0 + qV(z,t) \tag{2.2}$$

where $H_0$ is the Hamiltonian in the absence of interactions, $V(z,t)$ is an external electrical potential, and q is the electric charge. For zero mass electrons the free field Hamiltonian is given as,

$$H_0 = -i\sigma_3 \frac{\partial}{\partial z} \tag{2.3}$$

where $\sigma_3$ is the Pauli matrix with $\sigma_3 = \begin{pmatrix} 1 & 0 \\ 0 & -1 \end{pmatrix}$.

If the electrical potential is zero then the energy of a normalized state vector $|\Omega\rangle$ is given by,

$$\xi_0(t) = \langle\Omega|\int \hat{\psi}^\dagger(z,t) H_0 \hat{\psi}(z,t) dz |\Omega\rangle - \varepsilon_R \tag{2.4}$$

where $\varepsilon_R$ is a renormalization constant which is normally defined so that the energy of the vacuum state is zero. Since this is the energy when the electric potential is zero we will sometimes refer to it as the free field energy. The question we want to address is whether or not there is a lower bound to the free field energy. The way we will determine this is as follows. At the initial time $t = 0$ the electric potential is zero and the system is assumed to be in the initial unperturbed state. In this initial unperturbed state the field operator is defined by $\hat{\psi}(z,0) = \hat{\psi}_0(z)$, where $\hat{\psi}_0(z)$ is discussed in the Appendix, and the state vector is given by $|\Omega\rangle$. The initial energy is $\xi_0(0)$. Next apply an electric potential and then remove it at some later time $t_f$ so that,

$$V = 0 \text{ for } t \leq 0; \quad V \neq 0 \text{ for } 0 < t < t_f; \quad V = 0 \text{ for } t \geq t_f \tag{2.5}$$

The field operator $\hat{\psi}(z,t)$ will evolve in time according to Eq. (2.1) with the initial condition $\hat{\psi}(z,0) = \hat{\psi}_0(z)$. This will, in general, result in a change in the energy. At $t_f$, where the applied potential has been set back to zero, the energy is given by $\xi_0(t_f)$. Therefore the change in energy from $t=0$, to the final time, $t_f$ is given by,

$$\Delta\xi_0(0 \to t_f) = \xi_0(t_f) - \xi_0(0) \tag{2.6}$$

It will be shown that $\Delta\xi_0(0 \to t_f)$ can be a negative number with an arbitrarily large magnitude. Therefore, there is no lower bound to $\xi_0(t_f)$.

In order to calculate $\Delta\xi_0(0 \to t_f)$ take the time derivative of (2.4) to obtain,

$$\frac{d\xi_0(t)}{dt} = \langle\Omega|\int\left(\frac{\partial\hat{\psi}^\dagger(z,t)}{\partial t}H_0\hat{\psi}(z,t) + \hat{\psi}^\dagger(z,t)H_0\frac{\partial\hat{\psi}(z,t)}{\partial t}\right)dz|\Omega\rangle \tag{2.7}$$

Use (2.1) along with (2.2) and (2.3) to obtain,

$$\frac{d\xi_0(t)}{dt} = \langle\Omega|\int\left[\begin{pmatrix}i\dfrac{\partial\hat{\psi}^\dagger(z,t)}{\partial z}\dfrac{\partial\hat{\psi}(z,t)}{\partial z} + qV(z,t)\hat{\psi}^\dagger(z,t)\sigma_3\dfrac{\partial\hat{\psi}(z,t)}{\partial z}\end{pmatrix} + \begin{pmatrix}i\hat{\psi}^\dagger(z,t)\dfrac{\partial^2\hat{\psi}(z,t)}{\partial z^2} - q\hat{\psi}^\dagger(z,t)\sigma_3\dfrac{\partial}{\partial z}(V(z,t)\hat{\psi}(z,t))\end{pmatrix}\right]dz|\Omega\rangle \tag{2.8}$$

Integrate by parts to obtain,

$$\frac{d\xi_0(t)}{dt} = q\int\left(V(z,t)\frac{\partial}{\partial z}\langle\Omega|(\hat{\psi}^\dagger(z,t)\sigma_3\hat{\psi}(z,t))|\Omega\rangle\right)dz \tag{2.9}$$

Now, in order to continue this analysis, we need to solve the Dirac equation (2.1). The solution of (2.1) can be easily shown to be,

$$\hat{\psi}(z,t) = W(z,t)\hat{\psi}_0(z,t) \tag{2.10}$$

where $\hat{\psi}_0(z,t)$ is the solution to the free field equation,

$$i\frac{\partial\hat{\psi}_0(z,t)}{\partial t} = H_0\hat{\psi}_0(z,t) \tag{2.11}$$

and can be written as,

$$\hat{\psi}_0(z,t) = e^{-iH_0 t}\hat{\psi}_0(z) \tag{2.12}$$

The quantity $W(z,t)$ is given by,

$$W(z,t) = \begin{pmatrix} e^{-ic_1} & 0 \\ 0 & e^{-ic_2} \end{pmatrix} \quad (2.13)$$

where $c_1(z,t)$ and $c_2(z,t)$ satisfy the following differential equations,

$$\frac{\partial c_1}{\partial t} + \frac{\partial c_1}{\partial z} = qV \quad (2.14)$$

and,

$$\frac{\partial c_2}{\partial t} - \frac{\partial c_2}{\partial z} = qV \quad (2.15)$$

Use (2.10) in (2.9) to obtain,

$$\frac{d\xi_0(t)}{dt} = q\int \left( V(z,t) \frac{\partial}{\partial z} \langle \Omega | (\hat{\psi}_0^\dagger(z,t) W^\dagger(z,t) \sigma_3 W(z,t) \hat{\psi}_0(z,t)) | \Omega \rangle \right) dz \quad (2.16)$$

Use $W^\dagger(z,t)\sigma_3 W(z,t) = \sigma_3$ in the above to obtain,

$$\frac{d\xi_0(t)}{dt} = \int \left( V(z,t) \frac{\partial J_0(z,t)}{\partial z} \right) dz \quad (2.17)$$

where,

$$J_0(z,t) = q \langle \Omega | (\hat{\psi}_0^\dagger(z,t) \sigma_3 \hat{\psi}_0(z,t)) | \Omega \rangle \quad (2.18)$$

Integrate this from $t=0$ to $t_f$ to obtain,

$$\Delta\xi_0(0 \to t_f) = \int_0^{t_f} dt \int V(z,t) \frac{\partial J_0(z,t)}{\partial z} dz \quad (2.19)$$

Therefore at time $t_f$ the free field energy is given by,

$$\xi_0(t_f) = \Delta\xi_0(0 \to t_f) + \xi_0(0) \quad (2.20)$$

Note that in (2.19) the quantity $\partial J_0(z,t)/\partial z$ is independent of $V(z,t)$. This is evident from (2.18) and (2.12). Assume for the moment that $\partial J_0(z,t)/\partial z$ is non-zero. If this is the case then it is easy to show that we can always find a $V(z,t)$ which makes $\Delta\xi_0(0 \to t_f)$ a negative number with an arbitrarily large magnitude. For example let,

$$V(z,t) = -f \frac{\partial J_0(z,t)}{\partial z} \text{ for } 0 < t < t_f \quad (2.21)$$

where $f$ is a constant. Use this in (2.19) to obtain,

$$\Delta \xi_0 \left( 0 \rightarrow t_f \right) = -f \int_0^{t_f} dt \int \left| \frac{\partial J_0(z,t)}{\partial z} \right|^2 dz \qquad (2.22)$$

Now as $f \rightarrow +\infty$ it is evident that $\Delta \xi_0 \left( 0 \rightarrow t_f \right) \rightarrow -\infty$. This means that an arbitrarily large amount of energy has been extracted from the quantum state due to its interaction with the electric potential and that there is no lower bound to the final energy $\xi_0 \left( t_f \right)$.

**3. Discussion.**

This result may seem somewhat surprising because it contradicts the widely held assumption that there is a lower bound to the energy in quantum field theory. Therefore is worth carefully reviewing the assumptions that lead to these results. The first and main assumption is that the field operators obey the Dirac equation. Other than this we apply the normal rules of algebra and calculus to obtain (2.19). Note that we don't even use the commutation algebra so issues involving anomalous commutators are not a factor in these results.

Now to obtain the final result we assume that the quantum state has been set up so that $\partial J_0(z,t)/\partial z$ is non-zero. Now what is $J_0(z,t)$? $J_0(z,t)$ is the free field current expectation value of the normalized state vector $|\Omega\rangle$. That is, it is the current of the system in the absence of an electric potential. This is why it is independent of $V(z,t)$. Basically then we are assuming that a state vector $|\Omega\rangle$ exists where $\partial J_0(z,t)/\partial z$ is non-zero. Now how do we know that this is the case? It can be easily shown that when the field operator is expanded in the usual manner in terms of creation and annihilation operators that there are states that satisfy this condition. This is shown in the Appendix.

Therefore we have the following conclusion regarding our "very simple" field theory - if the field operator obeys the Dirac equation (2.1) and a state exists for which $\partial J_0(z,t)/\partial z$ is non-zero then there is no lower bound to the free field energy. This result is consistent with references [1-4] which show that there exist states with less energy than the vacuum in Dirac's hole theory and Ref. [5] in which it was shown that there exist states with less energy than the vacuum for quantum field theory in the Heisenberg picture.

**Appendix**

In the main body of this article we have not expressed the field operators in terms of creation and destruction operators because this was not necessary to achieve the main results. We will show that when this is done it is easy to shown that states exist where $\partial J_0(z,t)/\partial z$ is non-zero. Recall that the field operator $\hat{\psi}_0(z,t)$ must satisfy Eq. (2.11). In addition, it must satisfy the usual equal time anti-commutation relationship,

$$\{\hat{\psi}_{0,\alpha}(z,t),\hat{\psi}_{0,\beta}^{\dagger}(z',t)\} = \delta_{\alpha\beta}\delta(z-z') \tag{A.1}$$

Based on this we can write the field operator as,

$$\hat{\psi}_0(z,t) = \sum_p \left(\hat{b}_p \varphi_{1,p}^{(0)}(z,t) + \hat{d}_p^{\dagger} \varphi_{-1,p}^{(0)}(z,t)\right); \quad \hat{\psi}_0^{\dagger}(z,t) = \sum_r \left(\hat{b}_p^{\dagger} \varphi_{1,p}^{(0)\dagger}(z,t) + \hat{d}_p \varphi_{-1,p}^{(0)\dagger}(z,t)\right) \tag{A.2}$$

where the $\hat{b}_p$ ($\hat{b}_p^{\dagger}$) are the destruction(creation) operators for an electron associated with the state $\varphi_{1,p}^{(0)}(z,t)$ and the $\hat{d}_p$ ($\hat{d}_p^{\dagger}$) are the destruction(creation) operators for a positron associated with the state $\varphi_{-1,p}^{(0)}(z,t)$. They satisfy the anticommutator relationships,

$$\{\hat{d}_p,\hat{d}_q^{\dagger}\} = \delta_{pq}; \quad \{\hat{b}_p,\hat{b}_q^{\dagger}\} = \delta_{pq}; \text{ all other anti-commutators are zero} \tag{A.3}$$

The vacuum state $|0\rangle$ is defined by,

$$\hat{d}_p|0\rangle = \hat{b}_p|0\rangle = 0 \text{ and } \langle 0|\hat{d}_p^{\dagger} = \langle 0|\hat{b}_p^{\dagger} = 0 \text{ for all } p \tag{A.4}$$

Let $\varphi_{\lambda,p}^{(0)}(z)$ be the eigenfunctions of the free field Hamiltonian with energy eigenvalue $\varepsilon_{\lambda,p}^{(0)}$. They satisfy the relationship,

$$H_0 \varphi_{\lambda,p}^{(0)}(z) = \varepsilon_{\lambda,p}^{(0)} \varphi_{\lambda,p}^{(0)}(z) \tag{A.5}$$

where,

$$\varphi_{\lambda,p}^{(0)}(z) = \frac{1}{2\sqrt{L}} \begin{pmatrix} 1 + \dfrac{\lambda p}{|p|} \\ 1 - \dfrac{\lambda p}{|p|} \end{pmatrix} e^{ipz}; \quad \varepsilon_{\lambda,p}^{(0)} = \lambda|p| \tag{A.6}$$

and where $\lambda = \pm 1$ is the sign of the energy, $p$ is the momentum, and $L$ is the 1 dimensional integration volume. We assume periodic boundary conditions so that the momentum $p = 2\pi r/L$ where $r$ is an integer. According to the above definitions the quantities $\varphi_{-1,p}^{(0)}(z)$ are negative energy states with energy $\varepsilon_{-1,p}^{(0)} = -|p|$ and the quantities $\varphi_{+1,p}^{(0)}(z)$ are positive energy states with energy $\varepsilon_{+1,p}^{(0)} = |p|$.

The $\varphi_{\lambda,p}^{(0)}(z)$ form an orthonormal basis set and satisfy,

$$\int \varphi_{\lambda,p}^{(0)\dagger}(z)\varphi_{\lambda',p'}^{(0)}(z)\,dz = \delta_{\lambda\lambda'}\delta_{pp'} \tag{A.7}$$

where integration from $-L/2$ to $+L/2$ is implied. If the electric potential is zero then the $\varphi_{\lambda,p}^{(0)}(z)$ evolve in time according to,

$$\varphi_{\lambda,p}^{(0)}(z,t) = e^{-iH_0 t}\varphi_{\lambda,p}^{(0)}(z) = e^{-i\lambda|p|t}\varphi_{\lambda,p}^{(0)}(z) \tag{A.8}$$

Now consider the state vector,

$$|\Omega\rangle = \frac{1}{\sqrt{2}}\left(\hat{b}_{p,1}^\dagger + \hat{b}_{q,1}^\dagger\right)|0\rangle \tag{A.9}$$

where both $p$ and $q$ are positive numbers. Referring to the definitions in Section II we can show that the current expectation value of this state is,

$$J_0(z,t) = \frac{1}{2L}\left(1 + \cos\left((p-q)(z-t)\right)\right) \tag{A.10}$$

It is obvious, then, that $\partial J_0(z,t)/\partial z$ is non-zero.